\documentclass[%
 aip,
 amsmath,amssymb,
 reprint,%
]{revtex4-1}

\usepackage{graphicx}
\usepackage{dcolumn}
\usepackage{bm}

\usepackage[utf8]{inputenc}
\usepackage[T1]{fontenc}
\usepackage{mathptmx}
\usepackage{etoolbox}

\makeatletter
\def\@email#1#2{%
 \endgroup
 \patchcmd{\titleblock@produce}
  {\frontmatter@RRAPformat}
  {\frontmatter@RRAPformat{\produce@RRAP{*#1\href{mailto:#2}{#2}}}\frontmatter@RRAPformat}
  {}{}
}%
\makeatother
\begin{document}

\preprint{AIP/123-QED}

\title{Event-Based Imaging of Levitated Microparticles}
\author{Yugang Ren}
\address{ Joint first authors}
\affiliation{ 
Department of Physics, King's College London, Strand, London, WC2R 2LS, UK.
}%
\author{Enrique Benedetto}
\address{ Joint first authors}
\affiliation{ 
Department of Physics, King's College London, Strand, London, WC2R 2LS, UK.
}%
\author{Harry Borrill}
\affiliation{ 
Department of Physics, King's College London, Strand, London, WC2R 2LS, UK.
}%
\author{Yelizaveta Savchuk}
\affiliation{ 
Department of Physics, King's College London, Strand, London, WC2R 2LS, UK.
}%
\author{Molly Message}
\affiliation{ 
Department of Physics, King's College London, Strand, London, WC2R 2LS, UK.
}%
\author{Katie O'Flynn}
\affiliation{ 
Department of Physics, King's College London, Strand, London, WC2R 2LS, UK.
}%
\author{Muddassar Rashid}
\affiliation{ 
Department of Physics, King's College London, Strand, London, WC2R 2LS, UK.
}%

\author{James Millen}%
 \email{james.millen@kcl.ac.uk}
\affiliation{ 
Department of Physics, King's College London, Strand, London, WC2R 2LS, UK.
}%
\affiliation{%
London Centre for Nanotechnology, Department of Physics, King's College London, Strand, London, WC2R 2LS, UK.
}%

\begin{abstract}
\textbf{ABSTRACT}\\
\noindent Event-based imaging is a neuromorphic detection technique whereby an array of pixels detects a positive or negative change in light intensity at each pixel, and is hence particularly well suited to detecting motion. As compared to standard camera technology, an event-based camera reduces redundancy by not detecting regions of the image where there is no motion, allowing increased frame-rates without compromising on field-of-view. Here, we apply event-based imaging to detect the motion of a microparticle levitated under vacuum conditions, which greatly facilitates the study of nanothermodynamics and enables the independent detection and control of arrays of many particles. 
\end{abstract}

\maketitle

When unravelling the underlying physics of particles interacting with external forces, or of interacting multi-particle systems, object tracking is key. One must consider a range of detection metrics such as field-of-view, resolution, latency, sensitivity, bandwidth, signal-to-noise ratio (SNR) and the ability to detect multiple objects. 

In this work, we consider tracking the motion of microparticles levitated under vacuum conditions by optical, electrical or magnetic fields \cite{millen2020a, gonzalezballestro2022}. Such systems are of interest for studies of fundamental quantum science \cite{millen2020b}, nano-thermodynamics \cite{millen2018} and advanced sensing \cite{rademacher2020, moore2021}.

When working with particles optically trapped in liquid, it is sufficient to use standard CMOS or CCD cameras to track their motion, since the viscous damping provided by the liquid reduces dynamical timescales to a level suitable for camera frame-rates. However, once objects are levitated in vacuum, their motion is underdamped and faster tracking is required \cite{li2010}. Although this is possible at the 100\,kHz level with high-speed CMOS cameras \cite{svak21}, this requires significant reduction in the sensor resolution. Hence, particle tracking is usually performed using photodiodes, balanced photodetectors or quadrant photodetectors. Whilst these devices are fast, even tracking at GHz rates \cite{li2018,reimann2018}, they have limited field-of-view, restricting tracking to scales not much larger than the optical wavelength. This can limit studies of nanothermodynamics or nonlinear motion where particles may explore large regions of space \cite{millen2020b}.

\begin{figure}[t]
\includegraphics{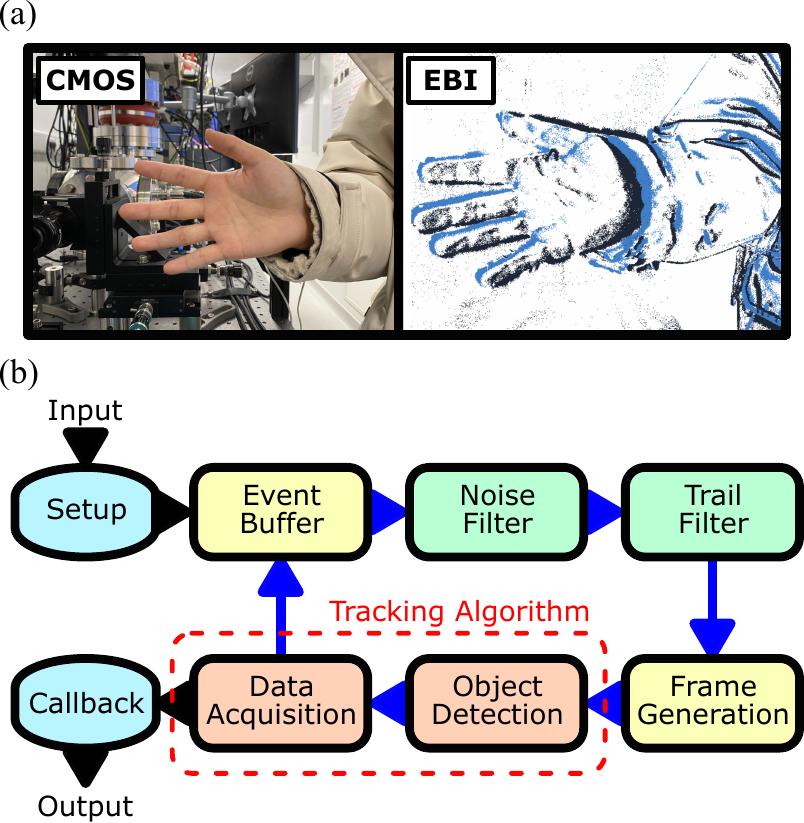}
\caption{(a) Left image is taken with a conventional CMOS camera, right image is taken by a camera using event-based imaging (EBI). The black and blue pixels in the image correspond to negative / positive changes in light intensity, respectively. (b) Information pipeline from the point of data capture in an event-based camera (EBC).}
\label{fig:figure1}
\end{figure}

Additionally, a key technique when working with levitated objects is the application of real-time feedback onto their motion. This has enabled cooling to the quantum ground state of a harmonic potential \cite{magrini2021}, but more generally is essential for stabilization under vacuum conditions. Balanced photodetectors are the standard tool to realise real-time feedback \cite{gieseler2012}, although recent work employing powerful graphics cards with a limited number of pixels \cite{bullier2019} or on-board microprocessors \cite{minowa2022} has enabled feedback control of sub-500\,Hz oscillators via CMOS camera detection. We also note that cameras have excellent SNR and allow super-resolution detection \cite{bullier2019}, allowing one to minimize the amount of light required to detect levitated particles avoiding absorption \cite{millen2014} and photon recoil heating \cite{jain2016}.

Finally, the prospect of levitating systems of multiple interacting particles has emerged \cite{gonzalezballestro2022}, for distributed sensing \cite{gadi2022} or generation of entanglement \cite{rudolph2022}. Single photodetectors can only track single particles, whereas cameras are well suited to multi-particle detection.

Conventional, CMOS/CCD based cameras work using a specified region of interest (ROI) or the whole pixel array to capture light from a scene. An alternative approach is that of event-based imaging (EBI), where pixels work independently of each other, triggering only when the change in light intensity is above a preset threshold \cite{gallego2022}. This enables a dynamic ROI, thus enabling decreased informational load compared to conventional cameras.

In this paper, we apply EBI for detecting the motion of microparticles levitated in vacuum. This imaging technique offers the potential for tracking single and multiple objects with high bandwidth, whilst an integrated tracking algorithm provides the real-time position of each object for use in feedback for state control.

Conventional cameras, such as CMOS/CCD, capture continuous movement as a sequence of still images (frames) formed from every pixel of the sensor. As a result, stationary elements are unnecessarily replicated, while moving elements are under-sampled \cite{Prophesee}.

EBI provides an approach to image acquisition by only capturing changes in images through the detection of modifications in light intensity on each pixel \cite{gallego2022}. Pixels in these neuromorphic sensors (sensors that try to mimic the neural structure of the brain) are completely independent. Each one of them contains a contrast detector (CD) that continuously tracks photocurrents. When the variation of a photocurrent crosses a threshold, the CD triggers a contrast detection event, which represents a relative increase (positive) or decrease (negative) in light intensity. It then initiates the measurement of a new value, as outlined in fig.\ \ref{fig:figure1}(b). Pixels which do not observe changes in light intensities that exceed the threshold do not produce output.

This can be seen in fig.\ \ref{fig:figure1}(a) where two images obtained with different cameras are shown. The event-based image consists of only three colours: white pixels indicate no change in light intensity across the threshold; blue and black pixels represent positive and negative changes in light intensity across the threshold, respectively. Only the blue and black pixels are transmitted as data. By comparing both sets of images in fig.\ \ref{fig:figure1}(a), it is clear that the amount of data transferred is suppressed with EBI.

In conventional cameras the bandwidth of the communication link is usually a constraint whenever higher acquisition rates are needed to track rapid movement. This leads users to reduce the ROI of the sensor to decrease the amount of data per frame \cite{bullier2019}. Due to the suppression of data redundancy, and the fact that EBI sensors have an effective pixel depth of 1-bit (an event is either detected or not), the data volume transmitted is considerably reduced, allowing acquisition rates over 1\,GHz \cite{gallego2022}.

Figure\ \ref{fig:figure1}(b) shows the detailed process of object tracking based on EBI. In an event-based camera (EBC), the EBI sensor is packaged with hardware which performs object tracking, such that the output of the EBC lists the position of each detected object as a function of time. All of the input information related to the camera sensor is read, and in the setup period camera sensor parameters are tuned. An event producer, which is contained in the event buffer, is used to generate a stream of events. For each stream of events, a noise filter is applied to pick up events in the neighboring 8 pixels during a certain time. A trail filter then accepts an event if the last event is detected at the same position within an accumulation time. All of the data collected by the sensor pixels generates a frame and a proprietary generic tracking algorithm (GTA) analyzes these frames to extract detected objects and associate data to previous frames. The deployed tracking algorithm limits the effective frame-rate of the EBC as compared to the read-out rate of the EBI sensor. When the detected object is recognized, there is a trigger in callback and the motional information of detected particle is obtained from output.

\begin{figure}[t]
\includegraphics{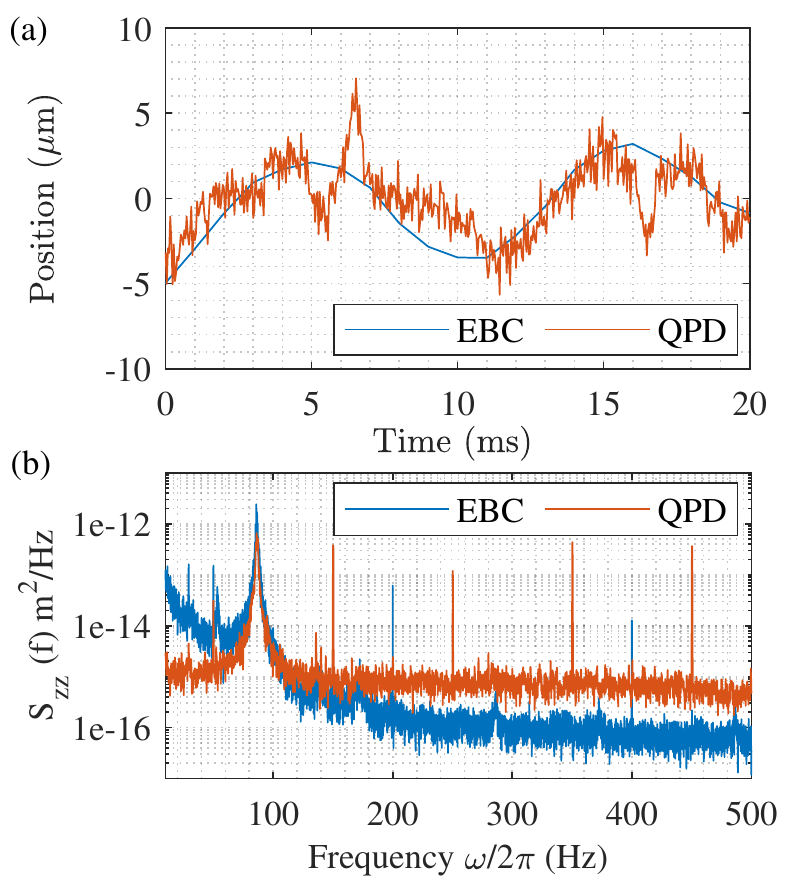}
\caption{(a) Time domain motion of one degree-of-freedom of a levitated microparticle, obtained via the GTA of an EBC (blue) and via a QPD (red). (b) Corresponding PSDs, illustrating the varying noise characteristics of the two detection methods.}
\label{fig:figure2}
\end{figure}

Detection based on object tracking (rather than, for example, measuring the intensity of light) has been shown to allow sub-pixel resolution and low noise \cite{bullier2019}. The dynamic range of the EBI sensor used in this study \footnote{Prophesee PPS3MVCD, $640 \times 480$ pixels} is 120\,dB, which is high compared to standard CMOS / CCD ($\sim 70\,$dB) or EMCCD ($\sim 100\,$dB) sensors. It is hard to quantify the effect of shot noise or dark counts on the GTA, and hence make a direct comparison to photodiode-based tracking.

\begin{figure*}[t]
\includegraphics{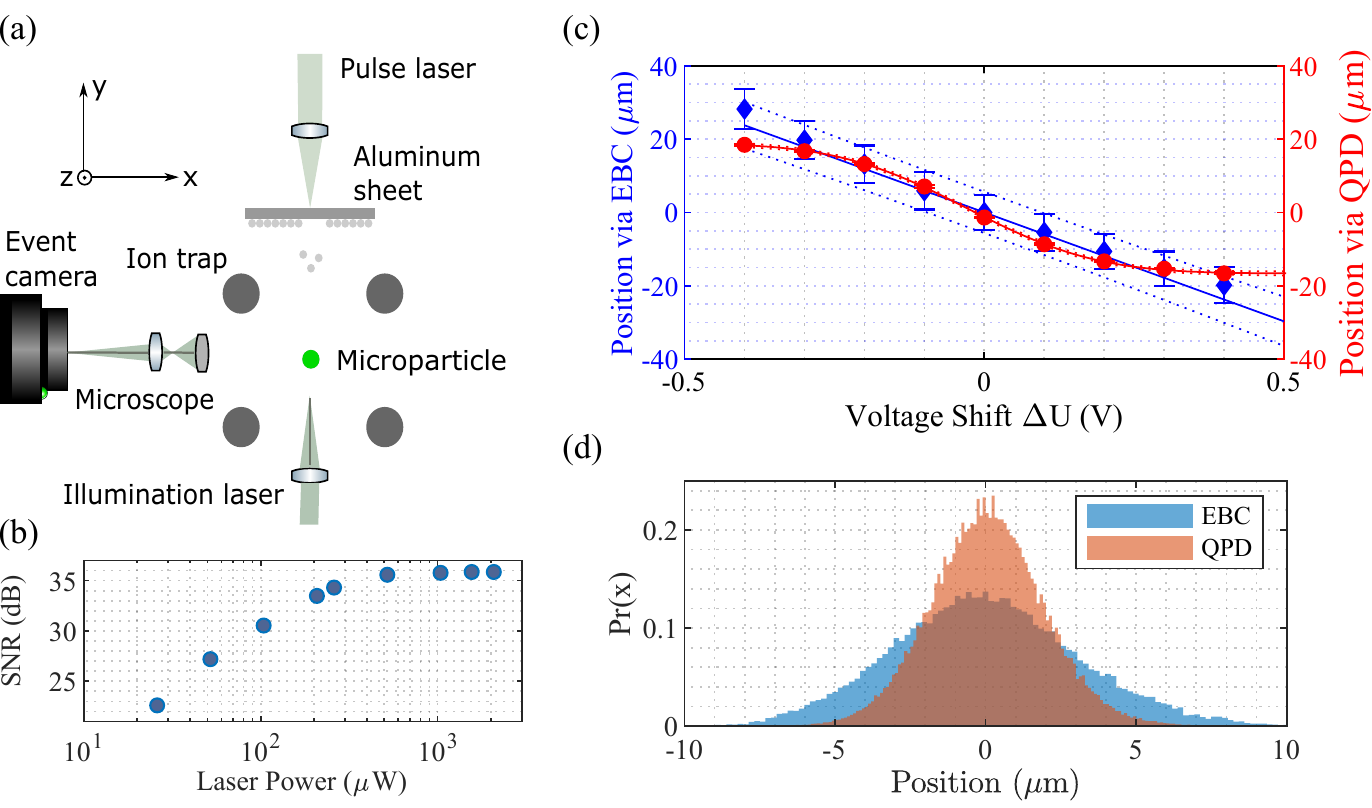}
\caption{(a) Schematic overview of experimental setup. Silica microspheres are levitated in a Paul trap, illuminated by a laser, and the scattered light is imaged onto an EBC or QPD. (b) EBC SNR at the motional frequency of the levitated microparticle, as a function of the maximal scattered laser power reaching the detector. (c) Average position of a levitated microparticle in response to a potential difference across the endcap electrodes. It is evident that the response of the EBC (blue) is linear across the full range of motion, whereas the QPD (red) responds non-linearly at large displacements. (d) Position probability histogram for the motion of a levitated microparticle, recorded by the EBC (blue) and QPD (red), where it is evident that the QPD cannot pick up large displacements.}
\label{fig:figure3}
\end{figure*}

We use an EBC\footnote{Prophesee Evaluation Kit Gen3M VGA CD 1.1} to track the motion of levitated microspheres. Charged silica microspheres of diameter 5\,$\mu$m and charge $Q$ are levitated in partial vacuum using a Paul trap, made with four 3mm-diameter rods and two 1mm-diameter endcap rods (not shown, aligned axially along the centre of the structure), as illustrated in fig.\ \ref{fig:figure3}(a). The microsphere is trapped using an oscillating electric field with frequency $\Omega_{RF} = 2\pi \times 800$\,rad\,s$^{-1}$ and amplitude $V_{RF}=750$\,V. The two endcap electrodes are held at -4V DC. The particles are loaded into the Paul trap using laser-induced acoustic desorption (LIAD) \cite{bykov2019, nikkhou2021} at a pressure of $2\times10^{-2}$ mbar.

An illumination laser (532\,nm) is focused onto the trapped particle, with a beam waist of $\sim 80\,\mu$m. The scattered light is collected by a microscope ($16 \times$ magnification) and directed onto the EBC for motion detection. In fig.\ \ref{fig:figure3}(b) we present the SNR at the motional frequency of the particle as recorded by the EBC, as we vary the power of the illumination laser (see Supplementary Information IV). We estimate the maximal scattered power reaching the EBC sensor, illustrating the excellent sensitivity of the detector.

Figure\ \ref{fig:figure2}(a) shows the output from the EBC compared to the output of a quadrant photodiode (QPD) \footnote{New Focus 2901 Quadrant Cell Photoreceiver}, when tracking the oscillatory motion of a microparticle using identical imaging systems. The GTA acts like an effective filter, removing high-frequency noise. Figure\ \ref{fig:figure2}(b) shows the corresponding derived power spectral densities (PSDs) of the particle motion. Like all balanced detection methods, the QPD minimizes $1/f$ noise at low frequencies, whereas the EBC has lower-noise performance at higher frequencies.

Regardless of the detection method used, the detector has to be calibrated. The motion of the particle is calibrated by applying a known potential difference $\Delta U$ across the endcap electrodes, causing the particle to move in the $z$-direction. The particle oscillates at the centre of the trap when $\Delta U = 0$\,V. If the voltage of one of the endcaps is modified from $U_0 \to U_1$, the resulting voltage difference $\Delta U=U_1-U_0$ exerts a force on the particle $\vec{F}=F_z \hat{z}$, where $\hat{z}$ denotes a unit vector along the $z$-axis, determined by:

\begin{equation}
    F_z = \frac{Q \left.\Delta U\right. }{d}\:,
\end{equation}

\noindent where $d$ is the distance between the endcap electrodes. We confirm this analytic model for our specific trap geometry in Supplementary Information \uppercase \expandafter {\romannumeral 2}. Since the Paul trap provides a harmonic pseudo-potential, the particle also experiences a linear restoring force:
\begin{equation}
    F_z = -k\left< z \right>\:,
\end{equation}

\noindent where $k$ is the trap stiffness and $\left< z \right>$ is the average position of the particle (assuming $\left<z\right>=0$ when $\Delta U=0$).

Noting that the measured $\langle z_m \rangle$ is in volts for the QPD, and pixels for the EBC, and related to the true value of $\langle z \rangle$ through a conversion factor $\gamma$, with units V/m and pixels/m, respectively, we can equate these two equations. Considering further that $k=m\omega^2_z$ for a harmonic oscillator of mass $m$ and oscillation frequency $\omega_z$, then:

\begin{equation}
    \left< z_m \right> = -\gamma\frac{Q}{m}\frac{\left.\Delta U\right.}{d \omega^2_z}\:.
    \label{eqn:real_displacement}
\end{equation}

By measuring the mean displacement of the particle as $\Delta U$ is varied, we can determine the conversion factor $\gamma$ for each detector. The oscillation frequency $\omega_z$ can be obtained from the power spectral density (fig.\ \ref{fig:figure2}(b)) and $Q/m$ obtained by solving the Mathieu equations (see Supplementary Information \uppercase \expandafter {\romannumeral 1}). In our system we typically trap particles of charge $Q = 2\times10^4\,e$.

The calibrated particle position as a response to a potential difference can be seen in fig.~\ref{fig:figure3}(c). It is apparent that as $\left|\Delta U\right|$ becomes large, the response of the QPD becomes nonlinear, whereas eqn.~(\ref{eqn:real_displacement}) predicts a linear response. This can be somewhat mitigated through a nonlinear calibration (see Supplementary Information \uppercase \expandafter {\romannumeral 2}), but limits the field-of-view of the QPD. On the other hand, the EBC has a linear response to the particle displacement across the full range. Figure~\ref{fig:figure3}(d) shows a histogram of the equilibrium motion of the trapped particle, again showing that large displacements are missing when using the QPD.  Hence, we can track both small (i.e. the oscillatory motion) and large displacements using EBI. 

Levitated particles provide an ideal system  for probing stochastic thermodynamics \cite{millen2018}. This is in part due to the characteristic energy of the system being comparable to that of thermal fluctuations of the bath, which enable levitated systems to be highly sensitive to surrounding fluctuations. The coupling to the bath is characterized by the ratio of its oscillation frequency $\omega_z$ to its momentum damping rate $\Gamma$, yielding overdamped ($\Gamma \gg \omega_z$) and underdamped ($\Gamma \ll \omega_z$) regimes. In the overdamped regime, trapped particles have been used for studying heat engines \cite{martinez2017colloidal}, non-thermal baths \cite{wei2010thermal}, and for testing Landauer's principle \cite{martinez2015stochastic}. In the underdamped regime, the studies in stochastic thermodynamics have been extended to observing Kramer's turnover \cite{rondin2017direct}, studying non-equilibrium Landauer's principle \cite{jun2014high} and testing fluctuation-dissipation theorems \cite{hoang2018experimental}.

Often when such systems exhibit non-equilibrium dynamics, they go beyond the linear detection regime of photodiode-based detection systems, and CMOS cameras would need a large ROI to capture the dynamics, at the expense of bandwidth. 

\begin{figure}[t]
\includegraphics{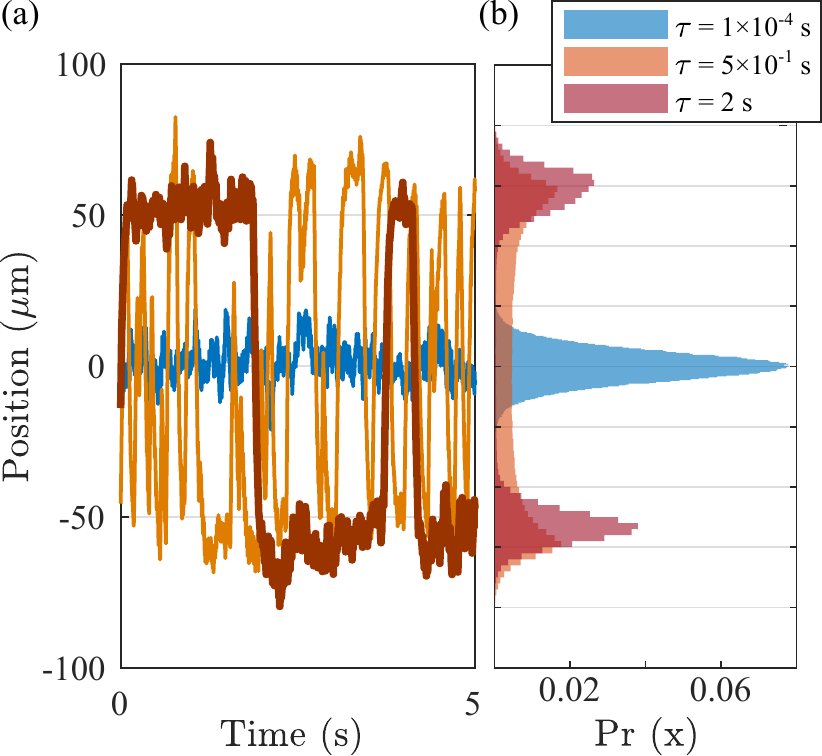}
\caption{(a) Particle motion, as tracked by an EBC, when the particle is driven to make large jumps in position. Different coloured lines indicate different jump time-constants $\tau$. (b) Position probability distributions for different jump time-constants. As predicted, when $\tau = 1\times10^{-4}$s, the distribution is Gaussian, and for slower jump time-constants the distribution is bimodal.}
\label{fig:figure4}
\end{figure}

EBI employs a dynamic ROI, based on only triggered pixels. To characterise the capability of EBI, and the GTA of the EBC, we cause random jumps in the particles' position with varying time intervals, $\tau$  (see Supplementary Information \uppercase \expandafter {\romannumeral 3}). These random jumps, as seen in fig.~\ref{fig:figure4}(a), are driven by voltage changes applied to the endcap electrodes which follow telegraph noise statistics (see Supplementary Information \uppercase \expandafter {\romannumeral 3}) distributed about a mean waiting time $\bar{\tau}$. For fast switching times  relative to the gas damping rate $\Gamma_0$, the position probability distribution of the particle is Gaussian, and for slow switching times it is bimodal,  as shown in fig.~\ref{fig:figure4}(b). Comparing to fig.~\ref{fig:figure3}(c), the particle's motion would go deep into the nonlinear range of the QPD. However, the EBC has no such limitation, and position shifts greater than $100\,\mu$m are tracked, without compromising on position sensitivity, which is approximately $30\,$nm\,Hz$^{-1/2}$ in this work for both devices, as can be extracted from fig.~\ref{fig:figure2}(b). For the QPD to track such a range of motion, one would have to use a lower magnification imaging system, with a corresponding reduction in position sensitivity.

The factor limiting the bandwidth of our EBC is the GTA. For shifts in position that are in quick succession and large in displacement, the GTA lags behind or misses consecutive shifts. This is illustrated in fig.~\ref{fig:figure4}(a), where the GTA of the EBC faithfully tracks the particle position when $\bar{\tau}>1\,$s. For shorter mean waiting times, $\bar{\tau}<500\,$ms the GTA struggles to track the jumps, as evidenced by apparent spikes in the time domain signal. The GTA doesn't represent a true bandwidth limit to EBI and advantageously, as shown in fig.~\ref{fig:figure4}(b), our EBC can simultaneously track large displacements and the smaller oscillations of the particle about its equilibrium position.

We extend our study to multi-particle tracking using EBI. The ability to track arrays of particles would enable the study of quantum correlations \cite{kotler2021direct, brandao2021coherent, mercier2021quantum, roque2012quantum}, non-hermitian systems \cite{qi2020topological}, and the detection of dark matter \cite{carney2021mechanical,moore2021}, vacuum friction \cite{zhao2012rotational} and differential force-sensing \cite{rudolph2022}. To date, few have experimentally explored multi-particle physics, with only two particles trapped to demonstrate state-swapping and sympathetic cooling \cite{penny2021sympathetic}, dipole-dipole \cite{rieser2022observation, arita2018optical} and coulomb-coulomb \cite{slezak2019microsphere} interactions, and cold damping and state control \cite{vijayan2022scalable}. 

Motional detection of individual particles in a multiple particle trap is a non-trivial problem. A single focused light beam can carry motional information of two particles \cite{praveen2016two}, or an additional beam can be added for detection of the second particle \cite{rieser2022observation}. Going beyond two particles in this way would require a complex and non-scalable optical setup. Within the context of optical tweezers, there are numerous approaches most suited to tracking multiple particles, using high-speed cameras \cite{gibson2008measuring} and multiple-beams in conjunction with a QPD \cite{ott2014simultaneous}, but these detection methods face the same limitations outlined earlier in this paper. 

\begin{figure}
\includegraphics{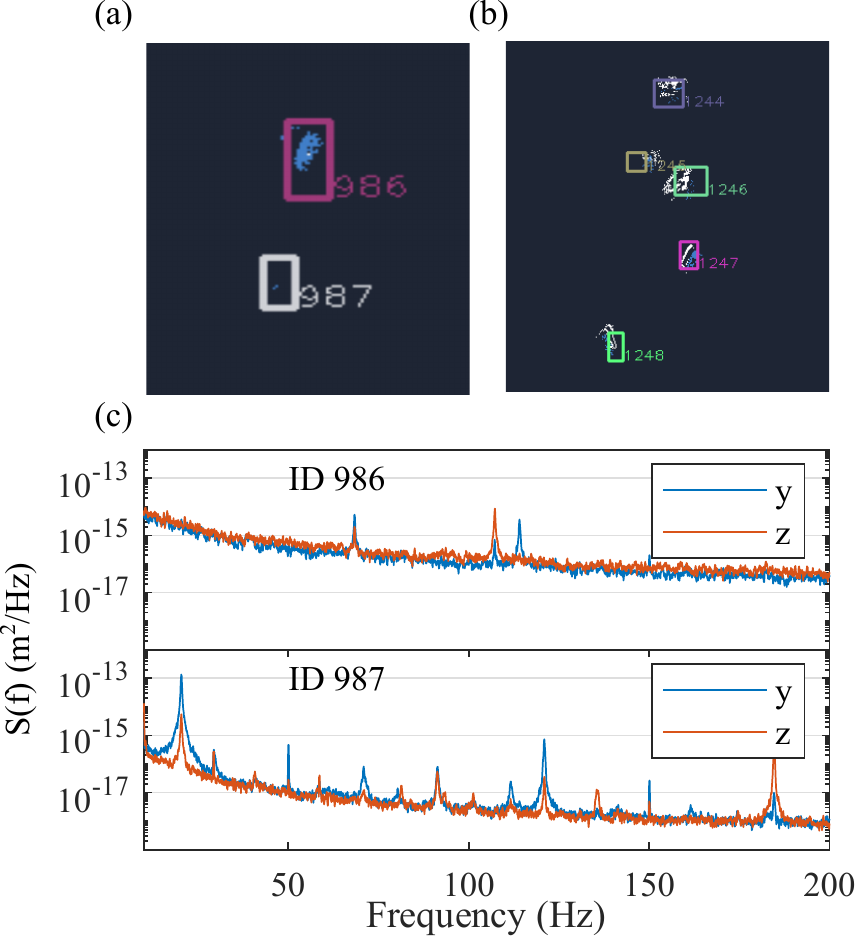}
\caption{(a) EBC image of two microspheres of $5\,\mu$m diameter trapped in a Paul trap. The image shows the bounding boxes which the GTA uses to track the two-dimensional position of the particles, labelled "986" and "987". (b) EBC image illustrating tracking of five particles. (c) PSDs reconstructed from the output of the GTA for the particles in (a).}
\label{fig:figure5}
\end{figure}

Figure~\ref{fig:figure5}(a) shows an image of two microspheres of $5\,\mu$m diameter trapped in a Paul trap, captured on our EBC. The image shows the bounding boxes which the GTA uses to track the two-dimensional position of the particles, labelled "986" and "987".

The levitated microspheres are tracked and their motional information separately reconstructed in fig.~\ref{fig:figure5}(c). We observe oscillation frequencies of both particles independently. For particle ID 986, the motional frequencies are $\{ \omega_x^{986} = 2\pi \times 68\,$rad\,s$^{-1}$, $\omega_y^{986} = 2\pi \times 114\,$rad\,s$^{-1}$, $\omega_z^{986} = 2\pi\times 108\,$rad\,s$^{-1}\}$, whilst particle ID 987 has frequencies $\{ \omega_x^{987} = 2\pi \times 20\,$rad\,s$^{-1}$, $\omega_y^{987} = 2\pi \times 122\,$rad\,s$^{-1}$, $\omega_z^{987} = 2\pi\times 184\,$rad\,s$^{-1}\}$. We observe all three centre-of-mass degrees-of-freedom due to imperfect alignment between the coordinate axis of the Paul trap and our imaging system. The additional frequency components in the spectrum for ID 987 could be other degrees-of-freedom (e.g. librational) or evidence of multi-particle collective modes, but that is beyond the scope of this study.

The EBC is not limited to tracking two particles, and in fig.~\ref{fig:figure5}(b) we show that the GTA has identified 5 separate particles. Due to varying $Q/m$, each particle has different resonant motional frequencies, and hence it is possible to individually excite them. We believe this opens the door to the study of a wide range of non-equilibrium phenomena.

In conclusion, we have shown that EBI is an interesting alternative to conventional detection schemes used for tracking levitated particles. The key enabling feature of EBI is the low data transfer, which enables EBC to track multiple objects at higher speeds than conventional cameras, and when combined with natural low pixel latencies \cite{gallego2022} will allow the experimenter to implement real-time feedback for state control. In this study, the bandwidth of tracking is limited to 1\,kHz by the very general and proprietary GTA employed. The underlying dynamics of our system are well known, therefore a more precise filter, like an asynchronous Kalman filter \cite{wang2021asynchronous, afshar2020event} in which we can input the expected equation of motion, will enable faster and more accurate tracking. 

As compared to photodiode-based detection schemes, EBCs feature a dynamic ROI, enabling tracking over a wide field-of-view, with particular application in the study of non-equilibrium physics. We have demonstrated tracking over 100 micrometres whist retaining 30\,nm\,Hz$^{-1/2}$ resolution. Finally, we have introduced an imaging technique suitable for fast tracking of a large number of particles, reaching MHz rates with the application of tailored particle tracking algorithms.

\section*{Supplementary Information}
Supplementary Material is provided for detailed information about how to solve charge-mass ratio, conduct nonlinear calibration of QPD, generate random jumps and obtain signal-to-noise ratio.
\\

JM would like to thank Dr. John Dale for the inspiration to start this project. This research has been supported by the European Research Council (ERC) under the European Union’s Horizon 2020 research and innovation programme (Grant Agreement Nos. 803277 \& 957463) and by EPSRC New Investigator Award EP/S004777/1.

\section{Calibration: Solving for $Q/m$}

Considering that our system follows a Mathieu equation of motion then the frequency of the different degrees-of-freedom can be written as,

\begin{equation}
    \label{eqn:mathieu_oscillation}
    \omega_i \cong \frac{1}{2} \Omega_{RF} \sqrt{a_i+\frac{1}{2}q_i^2},
    \renewcommand{\theequation}{S\arabic{equation} }  
\end{equation}

\noindent where $\omega_i$ the harmonic oscillation frequency, $\Omega_{RF}$ is the driving RF frequency, and $a_i$, $q_i$ with $i = \{x,y,z\}$ are known as the stability parameters \cite{drewsen2000harmonic}.

Using the stability parameters we can write a general statement for all three axis:
\begin{equation}
    \omega^2_i = q_m^2 \left( \frac{V_{RF}^2}{2\Omega_{RF}^2 r^4} \alpha_i^2 \right) + q_m \left( \frac{2 U}{d^2}\right) \beta_i,
    \renewcommand{\theequation}{S\arabic{equation} }
\end{equation}

\noindent where we $q_m = Q/M$ is the charge-mass ratio, $V_{RF}$ is the RF voltage amplitude, $U$ is the DC voltage at the endcaps, $r$ is the distance between RF electrodes from the centre of the trap and $d$ is the distance between the endcap electrodes. The variables $\alpha_i$ and $\beta_i$ are geometric factors that are obtained through SIMION numerical simulation of the Paul trap used in the experiments (see Table~\ref{tab:geometricfactors}) for each axis. The $Q/m$ ratio can then be obtained by solving for $q_m$. 

\begin{table} [h]
\renewcommand{\thetable}{S\arabic{table}}
\caption{\label{tab:geometricfactors}Geometric factors obtained from SIMION numerical simulation of our Paul Trap.}
\begin{tabular}{|c||c|c|c| } 
     \hline
             & $x$ & $y$ & $z$\\
     \hline
     $\alpha$ & 2.32 & 3.74 & 6.37 \\ 
     \hline
     $\beta$ & -0.090 & -0.090 & -0.198  \\ 
     \hline
    \end{tabular}
    
\end{table}

\section{Calibration: Nonlinear Calibration with the QPD}

To calibrate the system and obtain a conversion function we apply an electric force, $F_z$ to the particle via a potential difference $\Delta U$ across the endcap electrodes: 

\begin{equation}
    F_z = \frac{Q\Delta U}{d},
    \renewcommand{\theequation}{S\arabic{equation} }
\end{equation}

\noindent where $d$ is the spacing between the endcap electrodes, which is equal to spring force experienced in the trap: $m\omega^2_z \langle z \rangle$. When equated with the above equation we get a theoretical value for positional shift in metres for an applied voltage difference $\Delta U$: 

\begin{equation}\label{eq:x_m_thereotical}
    \langle z \rangle = \frac{Q}{m}\frac{\Delta U}{\omega_z^2 d}, 
    \renewcommand{\theequation}{S\arabic{equation} }
\end{equation}

\begin{figure}
\includegraphics{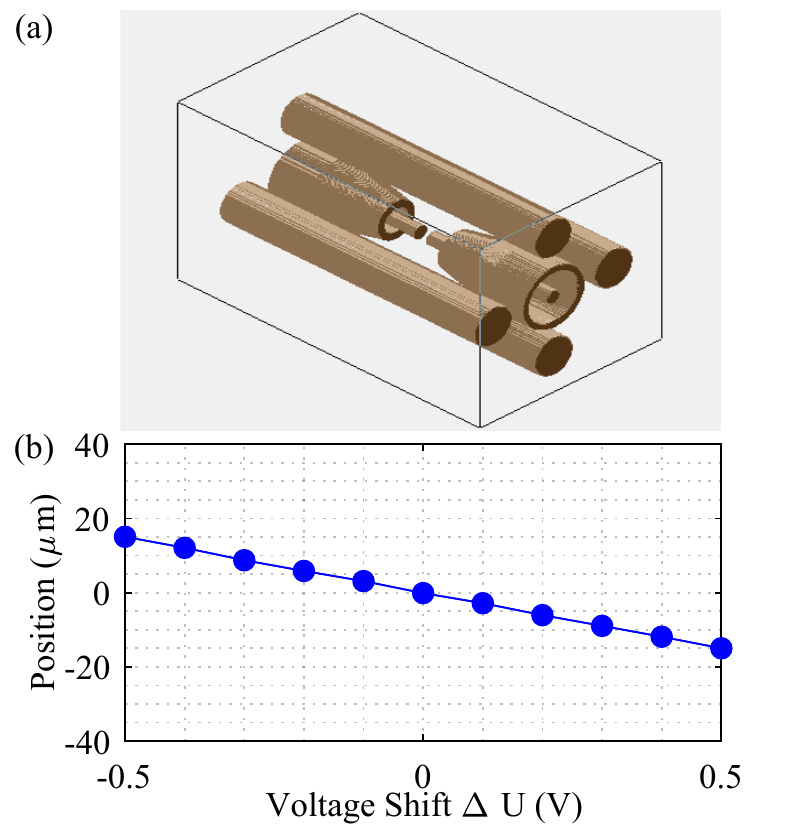}
\renewcommand{\figurename}{FIG.}
\renewcommand{\thefigure}{S\arabic{figure}}
\caption{\label {fig:simulation}  (a) Paul trap geometry in SIMION simulation environment which is set with the same size of experimental setup. (b) The mean position variations of the trapped particle with different voltage differences of two endcap rods, where a clear linear relationship can be seen.  }
\end{figure}

To further verify eqn.~S\ref{eq:x_m_thereotical}, simulations based on SIMION are conducted. Figure~\ref{fig:simulation}(a) shows a trap geometry in the simulation environment which is with the same size as our experimental system. Figure~\ref{fig:simulation}(b) shows that when the voltage of one endcap is tuned by -0.5V to 0.5V, related mean positions of the trapped particle linearly changes, verifiying the analytical expression used in the manuscript

\begin{figure}
\includegraphics{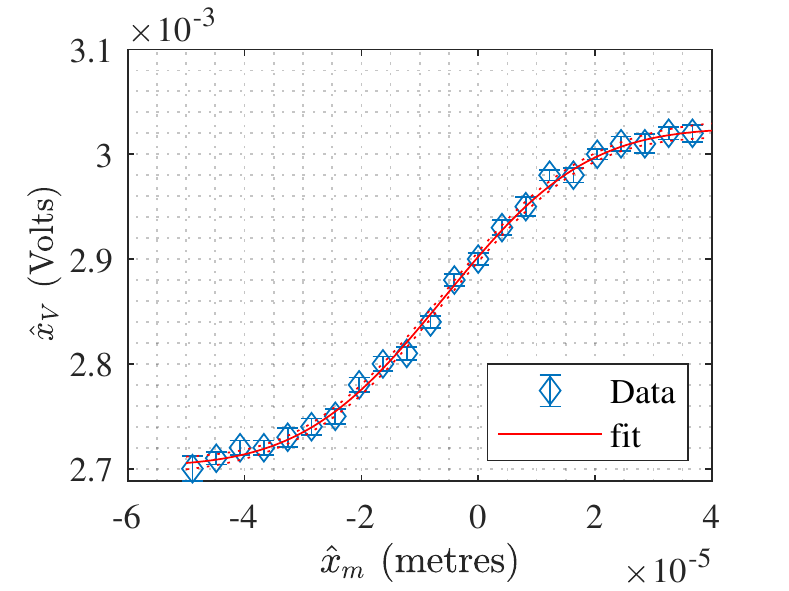}
\renewcommand{\figurename}{FIG.}
\renewcommand{\thefigure}{S\arabic{figure}}
\caption{\label {fig:NonLinFitting}  Error function fitted to the measured mean position in volts, $\langle x_V \rangle$ of the particle when a potential difference $\Delta U$ is applied across the endcaps. The measured position $\langle x_V \rangle$ is converted to the actual position $\langle x_m \rangle = \frac{qV}{m\omega^2 d}$. The inverse of the error function then converts Volts into metres}
\end{figure}

The quadrant photodetector (QPD) has a nonlinear response to the movement of an image if the image deviates too far from the centre, i.e. if the particle moves too far from the centre of the trap. We can fit an error function to the mean measured position of the particle (recorded in Volts, $\langle x_V \rangle$), with the expected actual position $\langle x_m \rangle$ based on the amount of applied potential difference, eqn.~S\ref{eq:x_m_thereotical}. 

\begin{equation}
    \langle x_V \rangle = a+b*\rm erf(\it \frac{\langle x_m \rangle_m-c}{d}),
    \renewcommand{\theequation}{S\arabic{equation} }
\end{equation}

\noindent where $(a,b,c,g)$ are fitting constants obtained by fitting the above equation to the data as shown in fig.~\ref{fig:NonLinFitting}. The inverse of this error function then enables us to convert the positional information in volts $\langle x_V \rangle$ to metres $\langle x_m \rangle$.

\begin{equation}
    \langle x_m \rangle = g\,\rm{erf}^{-1}{\it \left( \frac{\langle x_V \rangle-a}{b} \right) + c}.
    \renewcommand{\theequation}{S\arabic{equation} }
\end{equation}

This enables us to extend the detection range of the QPD by a few microns, but not by the tens of microns offered by the event-based camera.

\section{Generating Random Jumps}

Random jumps in position of the particle are implemented by applying telegraph noise statistics to the particle. To achieve this, we generate random numbers normally distributed, which then are multiplied by a mean waiting time $\bar{\tau}$. The waiting time is used to delay a voltage applied to the particle $\pm V_{\rm{tel}}$ via the endcap electrodes. This applied voltage is experienced by the trapped particle as an electric force,  $\pm\eta_{\rm{tel}} = \frac{Q}{m} V_{\rm{tel}}/(d)$. In this work, $V_{\rm{tel}} = \pm 1\,$V.

This applied force has noise statistics governed by:

\begin{equation}
    \langle \eta_{t} \eta_{t} \rangle = \eta_{\rm{tel}}^2 e^{(-2|t-t'|/\tau)}.
    \renewcommand{\theequation}{S\arabic{equation} }
\end{equation}
\\

\section{Detection Signal-to-noise ratio (SNR)}

We measure the SNR of the EBC at the resonant frequency of the particle motion $f_z$. To do this, we fit the following model to the noise floor of the detector, by analysing its power spectral density (PSD):
\begin{equation}
{\rm Noise}(f) = \log_{10} (\frac{a_{0}}{f}+\frac{b_{0}}{f^2}),
\renewcommand{\theequation}{S\arabic{equation} }
\end{equation}

where $a_0,b_0$ are fitting constants. We found that both terms were required to get a good fit. We fit the PSD of the particle motion with a standard model\cite{millen2018}, with the above noise model added:

\begin{equation}
{\rm Signal}(f) = \log_{10} (\frac{ac}{(b^2-f^2)^2+(cf)^2}+\frac{a_{0}}{f}+\frac{b_{0}}{f^2}),
\renewcommand{\theequation}{S\arabic{equation} }
\end{equation}

where $a,b,c$ are fitting constants.Therefore, the SNR at the motional frequency is obtained via ${\rm SNR}(f_z) = {\rm Signal}(f_z) - {\rm Noise}(f_z)$.

\section*{references}
\nocite{*}
\bibliography{aipsamp_3}

\end{document}